\documentclass[twocolumn]{aastex631}

\usepackage{amssymb}
\usepackage{amsmath,bm}
\usepackage{mathtools}
\usepackage{graphicx}
\usepackage{booktabs}
\usepackage{multirow}
\usepackage{xcolor}
\usepackage[normalem]{ulem}
\usepackage{hyperref}

\newcommand{\mr}{\mathrm}

\newcommand{\HPD}[3]{${ #1 }_{- #2}^{+ #3}$}

\renewcommand{\arraystretch}{1.5}

\renewcommand{\sout}{\bgroup \color{red} \ULdepth=-.5ex \ULset}

\hypersetup{linkcolor=blue, urlcolor=blue, citecolor=blue, anchorcolor=black}

\begin{document}

\title{On the Possibility of a Strong First-Order Phase Transition in Neutron Stars}

\author[0009-0002-9399-5598]{Zheng Cao}
\affiliation{State Key Laboratory of Dark Matter Physics, Key Laboratory for Particle Astrophysics and Cosmology (MOE), and Shanghai Key Laboratory for
Particle Physics and Cosmology, School of Physics and Astronomy,
Shanghai Jiao Tong University, Shanghai 200240, China}
\affiliation{Tsung-Dao Lee Institute, Shanghai Jiao Tong University, Shanghai 201210, China}

\author[0000-0002-7444-0629]{Lie-Wen Chen}
\affiliation{State Key Laboratory of Dark Matter Physics, Key Laboratory for Particle Astrophysics and Cosmology (MOE), and Shanghai Key Laboratory for
Particle Physics and Cosmology, School of Physics and Astronomy,
Shanghai Jiao Tong University, Shanghai 200240, China}

\correspondingauthor{Lie-Wen Chen}
\email{lwchen@sjtu.edu.cn}

\begin{abstract}
Whether cold dense QCD matter undergoes a strong first-order phase transition remains an open question.
In nature, neutron stars provide the most direct probe of cold dense QCD matter.
Theoretically, chiral effective field theory constrains the equation of state of dense matter near nuclear saturation density, while perturbative QCD calculations constrain it at densities well beyond stable neutron-star interiors.
We perform Bayesian inference with non-parametric Gaussian-process equation of state for $\beta$-equilibrated neutron-star matter under the assumption with and without a strong first-order phase transition, using the tidal deformability from GW170817, the NICER mass--radius measurements of PSR~J0740$+$6620, PSR~J0030$+$0451, PSR~J0437$-$4715, PSR~J0614$-$3329, chiral effective field theory, and perturbative QCD.
Within this Bayesian inference framework, the model comparison mildly to moderately favors a strong first-order phase transition, with its onset most likely lying \emph{above} the central density of the most massive neutron star.
Such an onset reconciles the stiffness required to support massive neutron stars with the softening favored by perturbative QCD from asymptotically high density.
\end{abstract}

\keywords{Neutron stars (1108) --- Nuclear astrophysics (1129) --- Nuclear physics (2077)}

\section{Introduction}
\label{sec:intro}

The phase structure of strong interaction matter at several times nuclear saturation density,
$n_0\equiv 0.16\,\mr{fm}^{-3}$, remains a central open problem in nuclear physics and astrophysics.
At low baryon chemical potential and high temperature, lattice quantum chromodynamics (QCD) and ultrarelativistic heavy-ion experiments have established that the transition from hadronic matter to the quark-gluon plasma is an analytic crossover~\citep{Aoki:2006we,Cheng:2006qk,Aoki:2009sc,Borsanyi:2013bia,HotQCD:2014kol}.
At low temperature and high baryon density, however, first-principles lattice calculations are obstructed by the sign problem~\citep{Alford:1998sd,Hands:2007by}.
Whether cold dense QCD matter contains a strong first-order phase transition (FOPT), characterized here by a vanishing sound speed over a finite density interval as in a sharp Maxwell construction~\citep{Glendenning:1992vb}, is therefore still unknown from first principles.

Neutron stars provide the most direct observational access to cold dense matter.
Recent multimessenger measurements have substantially narrowed the uncertainties of equation of state (EOS) of cold dense $\beta$-equilibrated matter.
In particular, the gravitational wave signal GW170817 from the binary neutron-star merger constrains the tidal deformability of canonical-mass neutron stars~\citep{LIGOScientific:2017vwq,LIGOScientific:2018hze}, while the Neutron Star Interior Composition Explorer (NICER) provides simultaneous mass--radius measurements for several pulsars through pulse-profile modeling.
The NICER measurement of the high-mass pulsar PSR~J0740$+$6620~\citep{Miller:2021qha,Riley:2021pdl,Salmi:2024aum} yields a mass $M\simeq 2.08\,M_\odot$ that places a stringent lower bound on the neutron-star maximum mass, while its measured radius constrains the EOS at high density.
The lower-mass NICER sources PSR~J0030$+$0451~\citep{Miller:2019cac,Riley:2019yda,Vinciguerra:2023qxq}, PSR~J0437$-$4715~\citep{Choudhury:2024xbk,Reardon:2024rdv}, and PSR~J0614$-$3329~\citep{Mauviard:2025dmd} further constrain the neutron-star radii in the canonical-mass regime.
Together, current observations probe the EOS of neutron-star matter over baryon densities of roughly $3$--$8\,n_0$ reached inside stable neutron stars.

Theoretical calculations constrain the EOS in complementary density regimes.
At low density, chiral effective field theory (ChEFT) provides controlled calculations up to $\sim 1$--$2\,n_0$, with quantified many-body and truncation uncertainties~\citep{Tews:2012fj,Hebeler:2013nza,Lynn:2015jua,Drischler:2017wtt,Drischler:2020hwi,Keller:2022crb}.
At very high density, perturbative QCD (pQCD) provides a weak-coupling calculation of the EOS once the quark chemical potential is sufficiently large~\citep{Freedman:1976ub,Kurkela:2009gj,Kurkela:2016was,Gorda:2018gpy,Gorda:2021znl,Gorda:2023usm}.
Although pQCD thermodynamics becomes quantitatively reliable only at extremely high densities of order $40\,n_0$, causality and thermodynamic stability allow the high-density constraint to be propagated downward~\citep{Komoltsev:2021jzg}.
Recent work has shown that the pQCD speed of sound remains well converged down to about $25\,n_0$~\citep{Gorda:2023usm}, extending the validity range of pQCD toward lower densities.
Nevertheless, in the broad density range between the ChEFT and pQCD regimes, it is still unclear whether the EOS varies smoothly (no strong first-order phase transition, NPT) or contains an FOPT.

These observational and theoretical inputs have sparked extensive studies of the FOPT in cold dense QCD matter.
Most analyses~\citep{Alford:2013aca,Alvarez-Castillo:2017qki,Ayriyan:2017nby,Sieniawska:2018zzj,Montana:2018bkb,Han:2018mtj,
Christian:2019qer,Pang:2020ilf,Blacker:2020nlq,Annala:2019puf,Tang:2020koz,Tan:2021ahl,Tsaloukidis:2022rus,
Gorda:2022lsk,Brandes:2023hma,Takatsy:2023xzf,Essick:2023fso,Kumar:2023lhv,Annala:2023cwx,
Christian:2023hez,Zhou:2024yzy,Saha:2024vst,
Ayriyan:2025rub,Grundler:2025mcz,Verma:2025vkk,Ji:2025sfn,Li:2025obt,Lindblom:2025wme,Saha:2025don,
Hammond:2025kki,Ecker:2025vnb}, however, do not directly compare the FOPT and NPT hypotheses through Bayesian model selection.
Such a comparison is carried out in only a few studies~\citep{Pang:2021jta,Komoltsev:2024lcr,Huang:2025vfl,Tang:2025xib}, but the FOPT onset density they consider is generally restricted to densities reached in neutron-star interiors.
It is therefore worthwhile to test the FOPT hypothesis over a broader density range that spans the full ChEFT--pQCD window.

In this work, we perform Bayesian inference using a non-parametric Gaussian-process (GP) EOS for cold $\beta$-equilibrated neutron-star matter.
The inference combines the tidal deformability of GW170817, NICER mass--radius measurements of PSR~J0740$+$6620, PSR~J0030$+$0451, PSR~J0437$-$4715, and PSR~J0614$-$3329, ChEFT below $1.5\,n_0$, and pQCD constraints.
The GP EOS is terminated at $n_L=25\,n_0$ to span the full ChEFT--pQCD window, with a lower choice $n_L=12\,n_0$ examined for comparison.
Comparing the NPT hypothesis with the FOPT hypothesis in which $c_s^2=0$ over a finite density interval, we find mild to moderate evidence in favor of an FOPT over NPT, with the FOPT onset density most likely located above the central density of the maximum-mass neutron star.

This paper is organized as follows.
Section~\ref{sec:methods} describes the Gaussian-process EOS construction and the Bayesian framework combining GW170817, NICER, ChEFT, and pQCD constraints.
Section~\ref{sec:results} presents the inferred FOPT properties and their implications for the EOS, stellar observables, and the location of the transition relative to stable neutron-star interiors.
Section~\ref{sec:summary} summarizes the main conclusions.

\section{Methods}
\label{sec:methods}
\subsection{Physics-agnostic EOS}

To construct a physics-agnostic and non-parametric EOS of cold $\beta$-equilibrated neutron-star matter, we model its squared sound speed $c_s^2(n)$ at baryon number density $n$ with GP regression~\citep{Rasmussen:2006}.
Causality and thermodynamic stability bound $c_s^2$ to $[0,1]$, so we place the GP prior on the auxiliary variable $\phi(n)\equiv -\ln\!\left[1/c_s^2(n)-1\right]$ rather than on $c_s^2$ itself,
\begin{equation}
\label{eq:gp}
\phi(n)\sim \mathcal{GP}\!\left(\bar\phi,\,K(n,n')\right),\qquad \bar\phi\equiv -\ln\!\left[1/\bar c_s^2-1\right],
\end{equation}
with squared-exponential kernel $K(n,n')=\eta\exp[-(n-n')^2/(2\ell^2)]$.
The hyperprior distributions are~\citep{Gorda:2022jvk}
\begin{align}
\label{eq:hyperprior}
\ell &\sim \mathcal{N}(1.0\,n_0,(0.25\,n_0)^2), \nonumber\\
\eta &\sim \mathcal{N}(1.25,0.2^2), \\
\bar{c}_s^2 &\sim \mathcal{N}(0.5,0.25^2). \nonumber
\end{align}
The GP variable $\phi(n)$ is predicted over densities from a crust-matching density $n_\mathrm{cc}=0.5\,n_0$ up to a termination density $n_L$.
Since pQCD constrains a nonzero $c_s^2$ down to $25\,n_0$~\citep[e.g., Fig.~10 of][]{Komoltsev:2023zor}, we set $n_L=25\,n_0$ to span the full ChEFT--pQCD interval. We also examine $n_L=12\,n_0$ for comparison.

The GP of Eq.~\eqref{eq:gp} is conditioned on theoretical calculations at a set of training densities $\mathbf{n}_t=\mathcal{T}_\mr{ChEFT}\cup\mathcal{T}_\mr{pQCD}$, comprising a low-density ChEFT subset $\mathcal{T}_\mr{ChEFT}$ and a high-density pQCD subset $\mathcal{T}_\mr{pQCD}$.
The ChEFT subset is a grid aligned with the next-to-next-to-next-to-leading-order (N$^3$LO) ChEFT calculation of \citet{Keller:2022crb}, spanning $0.58\,n_0\le n\le 1.5\,n_0$.
At each $n_t\in\mathcal{T}_\mr{ChEFT}$, the mean $c_{s,t}^2$ is the midpoint of the softest and stiffest ChEFT sound speeds, and $\sigma_t$ is fixed so that the $90\%$ interval of $\mathcal{N}(c_{s,t}^2,\sigma_t^2)$ matches the ChEFT band.
The pQCD subset $\mathcal{T}_\mr{pQCD}=\{25\,n_0\}$ anchors the GP to the pQCD prediction at $25\,n_0$, where $c_s^2(25\,n_0)\sim\mathcal{N}(c_{s,t}^2,\sigma_t^2)$.
Following \citet{Komoltsev:2023zor}, we take $c_{s,t}^2=0.32\,c^2$ as the mean and $\sigma_t=0.05\,c^2$ as twice the standard deviation of the pQCD predictions at $25\,n_0$~\citep{Komoltsev:QCDcode}.

Each $c_s^2(n)$ realization is drawn from the GP conditioned on the training data.
The training inputs are mapped into the GP variable through $\boldsymbol{\phi}_t=\phi(\boldsymbol{c}_{s,t}^2)$, while propagating $\boldsymbol{\sigma}_t$ through $\phi$ gives the diagonal training covariance $\boldsymbol{\Sigma}_t=\mathrm{diag}\!\left\{\frac{1}{4}\left[\phi(\boldsymbol{c}_{s,t}^2+\boldsymbol{\sigma}_t)-\phi(\boldsymbol{c}_{s,t}^2-\boldsymbol{\sigma}_t)\right]^2\right\}$.
Conditioning the GP prior on $\boldsymbol{\phi}_t$ and $\boldsymbol{\Sigma}_t$ then yields the predicted values $\boldsymbol{\phi}_\star\equiv\phi(\mathbf{n}_\star)$ at prediction densities $\mathbf{n}_\star\in[n_\mathrm{cc},n_L]$, following a multivariate Gaussian,
\begin{equation}
p(\boldsymbol{\phi}_\star\mid\boldsymbol{\phi}_t,\boldsymbol{\Sigma}_t)=\mathcal{N}\!\left(\boldsymbol{\mu}_\star,\,\boldsymbol{\Sigma}_\star\right),
\end{equation}
with mean and covariance
\begin{align}
\boldsymbol{\mu}_\star &= \bar\phi + \mathbf{K}(\mathbf{n}_\star,\mathbf{n}_t)\,\mathbf{C}^{-1}\,(\boldsymbol{\phi}_t-\bar\phi),\\
\boldsymbol{\Sigma}_\star &= \mathbf{K}(\mathbf{n}_\star,\mathbf{n}_\star) - \mathbf{K}(\mathbf{n}_\star,\mathbf{n}_t)\,\mathbf{C}^{-1}\,\mathbf{K}(\mathbf{n}_t,\mathbf{n}_\star),
\end{align}
where $\mathbf{C}\equiv \mathbf{K}(\mathbf{n}_t,\mathbf{n}_t)+\boldsymbol{\Sigma}_t$.
Sampling $\boldsymbol{\phi}_\star$ from $p(\boldsymbol{\phi}_\star\mid\boldsymbol{\phi}_t,\boldsymbol{\Sigma}_t)$ and inverting the map $\phi$ yields $c_s^2(n)$.

The construction above corresponds to the NPT hypothesis, where the GP yields a smooth $c_s^2(n)$ throughout $[n_\mathrm{cc},n_L]$.
For the FOPT, we adopt a uniform-density prior for the phase-transition location by sampling two densities  uniformly over $[1.5\,n_0,\,n_L]$ and labeling the smaller and larger values as the phase-transition onset density $n_S$ and end density $n_E$, respectively.
In the Maxwell construction, the sound speed vanishes between them,
\begin{equation}
c_s^2(n)=0,\qquad n_S\le n\le n_E.
\end{equation}
We then draw two independent GP realizations over $[n_\mathrm{cc},n_L]$ and take the first on $[n_\mathrm{cc},n_S]$ and the second on $[n_E,n_L]$, leaving the sound speed on the two sides of the transition uncorrelated.

Once $c_s^2(n)$ is specified for $n \ge n_\mathrm{cc}$ under either hypothesis, the EOS follows from the thermodynamic relations
\begin{align}
\mu(n) &= \mu(n_\mathrm{cc})\exp\!\left[\int_{n_\mathrm{cc}}^{n}\!\mathrm{d}n'\,\frac{c_s^2(n')}{n'}\right], \label{eq:mu}\\
\varepsilon(n) &= \varepsilon(n_\mathrm{cc}) + \int_{n_\mathrm{cc}}^{n}\!\mathrm{d}n'\,\mu(n'), \label{eq:eps}\\
p(n) &= -\varepsilon(n)+\mu(n)\,n, \label{eq:p}
\end{align}
where $\mu$, $\varepsilon$, and $p$ denote the baryon chemical potential, energy density, and pressure, respectively.
For $n\le n_\mathrm{cc}$, we adopt the crust EOS~\citep{Baym:1971pw,Negele:1971vb}, which fixes the boundary values $\mu(n_\mathrm{cc})$ and $\varepsilon(n_\mathrm{cc})$ in Eqs.~\eqref{eq:mu} and \eqref{eq:eps}.

We solve the Tolman--Oppenheimer--Volkoff equations~\citep{Tolman:1939jz,Oppenheimer:1939ne} for each EOS sample to obtain the neutron-star mass--radius relation and the central baryon density $n_c$ of the most massive configuration.
Within the FOPT hypothesis, we further distinguish whether the transition occurs inside stable neutron stars.
Samples with $n_S<n_c$ are labeled FOPT-in, while samples with $n_S\ge n_c$ are labeled FOPT-out.
The dimensionless tidal deformability $\Lambda$ is obtained from the $\ell=2$ tidal Love number $k_2$~\citep{Hinderer:2007mb}.
Under the FOPT hypothesis, the energy-density discontinuity $\Delta\varepsilon$ at the transition radius $r_d$ shifts the metric perturbation variable $y\equiv r H'/H$ by $\Delta y=-4\pi r_d^3\,\Delta\varepsilon/[m(r_d)+4\pi r_d^3\,p(r_d)]$~\citep{Postnikov:2010yn}, which we incorporate when integrating $k_2$ across the transition.

An FOPT EOS may give rise to two stable branches in the mass--radius relation, allowing for twin-star configurations.
To detect this behavior, we solve the TOV equations on a grid of compact-star central pressures $ p_c $ spaced log-uniformly from $0.5\,\mr{MeV\,fm^{-3}}$ up to the pressure at the termination density $n_L$ (using $150$ points for $n_L=25\,n_0$ and $100$ points for $n_L=12\,n_0$), and identify a second stable branch whenever the mass $M(p_c)$ rises again over at least three consecutive solutions  after passing through a local minimum.
It should be emphasized that the Seidov condition~\citep{Seidov71} $\Delta\varepsilon_\mr{crit}/\varepsilon_\mr{t}=\tfrac{1}{2}+\tfrac{3}{2}\,p_\mr{t}/\varepsilon_\mr{t}$ (with $\varepsilon_\mr{t}$ and $p_\mr{t}$ the energy density and pressure at the transition), often invoked for an FOPT, does not directly determine the existence of a twin star.
In fact, stable second branches can arise for both $\Delta\varepsilon<\Delta\varepsilon_\mr{crit}$ and $\Delta\varepsilon>\Delta\varepsilon_\mr{crit}$, as illustrated by Figs.~2(b) and 2(d), respectively, of \citet{Alford:2013aca}.
We therefore classify twin stars directly from the stability of the computed sequence in the following.

\begin{deluxetable*}{ll}
\tablecaption{Likelihood terms used in the Bayesian inference. For each term, the second column lists the reference and data from which the likelihood is built, together with the hot-spot model (for the NICER mass--radius measurements) or the spin prior (for GW170817).
\label{tab:observational_samples}}
\tablehead{
\colhead{Likelihood} &
\colhead{Source}
}
\startdata
PSR~J0740$+$6620 &
Ref.~\citet{Salmi:2024aum,Salmi6620}; ST-U hot-spot model. \\
PSR~J0030$+$0451 &
Ref.~\citet{Vinciguerra:2023qxq,Vinciguerra0451}; ST$+$PDT hot-spot model. \\
PSR~J0437$-$4715 &
Ref.~\citet{Choudhury:2024xbk,Choudhury4715}; CST$+$PDT hot-spot model. \\
PSR~J0614$-$3329 &
Ref.~\citet{Mauviard:2025dmd,Lucien0614}; ST$+$PDT hot-spot model. \\
GW170817 &
Ref.~\citet{LIGOScientific:2017vwq,hernandez_toast_2020}; low-spin prior. \\
{\footnotesize Marginalized pQCD} &
Ref.~\citet{Komoltsev:2023zor,Komoltsev:2025marg}. \\
\enddata
\end{deluxetable*}

\subsection{Bayesian inference}

We combine the theoretical and observational information in a hierarchical Bayesian framework.
For data $\vec{d}$ and hypothesis $\mathcal{H}$, the posterior for the EOS parameters $\theta$ is~\citep{Thrane:2018qnx,HernandezVivanco:2020cyp}
\begin{equation}\label{eq:bayes}
p(\theta\,|\,\vec{d},\mathcal{H}) =
\frac{\prod_i \mathcal{L}(d_i\,|\,\theta,\mathcal{H})\,\pi(\theta\,|\,\mathcal{H})}
{\mathcal{Z}_{\mathcal{H}}(\vec{d})},
\end{equation}
where $i$ runs over independent constraints and $\pi(\theta\,|\,\mathcal{H})$ denotes the prior under hypothesis $\mathcal{H}$.
The evidence is
\begin{equation}
\mathcal{Z}_{\mathcal{H}}(\vec{d})\equiv
\int \prod_i \mathcal{L}(d_i\,|\,\theta,\mathcal{H})\,
\pi(\theta\,|\,\mathcal{H})\,\mathrm{d}\theta ,
\end{equation}
and it quantifies the support that the data give to the hypothesis.
We compare two hypotheses through the Bayes factor~\citep{Jeffreys:1939xee,Lee14}
\begin{equation}\label{eq:BF}
\mathcal{B}^{1}_{2}=\mathcal{Z}_{1}(\vec{d})/\mathcal{Z}_{2}(\vec{d}).
\end{equation}
On the Jeffreys scale~\citep{Jeffreys:1939xee,Trotta:2008qt,Lee14}, $\mathcal{B}^{1}_{2}$ in the ranges $[1,3]$, $[3,10]$, $[10,30]$, $[30,100]$, and $[100,\infty)$ corresponds to anecdotal, moderate, strong, very strong, and extreme evidence, respectively, in favor of hypothesis~$1$.

For each NICER mass--radius measurement, the likelihood is obtained by marginalizing a kernel density estimate, with the bandwidth set by Scott's rule~\citep{Scott:1992}, of the published mass--radius posterior over the neutron-star mass along the EOS-predicted curve $R(m,\theta)$.
The analysis includes PSR~J0740$+$6620~\citep{Salmi:2024aum,Salmi6620}, PSR~J0030$+$0451~\citep{Vinciguerra0451,Vinciguerra:2023qxq}, PSR~J0437$-$4715~\citep{Choudhury:2024xbk,Choudhury4715}, and PSR~J0614$-$3329~\citep{Mauviard:2025dmd,Lucien0614}.
The published posteriors for PSR~J0740$+$6620, PSR~J0437$-$4715, and PSR~J0614$-$3329 already include the corresponding radio-timing mass information, whereas PSR~J0030$+$0451 has no independent radio-timing mass prior.
For the tidal deformability from GW170817~\citep{LIGOScientific:2017vwq,LIGOScientific:2018hze}, we adopt the nuisance-marginalized likelihood of \citet{HernandezVivanco:2020cyp,hernandez_toast_2020} under the low-spin prior and marginalize over the component masses along the EOS-predicted curve $\Lambda(m,\theta)$.
Throughout the Bayesian analysis, we adopt a flat prior on the compact-star mass $m$.

The above astrophysical likelihoods constrain the EOS only within stable neutron stars ($n\lesssim 5$--$8\,n_0$).
To extend the constraint to higher densities, we adopt the marginalized pQCD likelihood of \citet{Komoltsev:2023zor, Komoltsev:2025marg}.
The likelihood is built from an auxiliary Gaussian process $\mathrm{GP_{aux}}$, anchored to the pQCD thermodynamic quantities at $n_\mr{pQCD}\simeq 40\,n_0$~\citep{Gorda:2021znl} and conditioned on the pQCD-predicted $c_s^2(n)$ over $25\,n_0\le n\le 40\,n_0$, where perturbative corrections remain small~\citep{Gorda:2023usm}.
For any $n_m\leq 25\,n_0$, marginalizing $\mathrm{GP_{aux}}$ over $[n_m,\,n_\mr{pQCD}]$ yields a joint distribution of $(p, \varepsilon)$ at $n_m$, which defines the marginalized pQCD likelihood at $n_m$.
We evaluate this likelihood directly at the EOS termination density $n_L$, i.e., $n_m=n_L$, thereby constraining the entire EOS below $n_L$.
Two caveats are in order. First, the pQCD likelihood constrains $(p,\varepsilon)$ at $n_L$, reflecting the integral of $c_s^2$ up to $n_L$ rather than $c_s^2(n_L)$ itself.
Second, $\mathrm{GP_{aux}}$ is smooth by construction and therefore does not model an FOPT between $n_L$ and $n_\mr{pQCD}$.
The likelihood terms, together with their references and data, are summarized in Table~\ref{tab:observational_samples}.
In summary, the likelihood terms listed in Table~\ref{tab:observational_samples}, together with the ChEFT sound-speed constraint at the low-density training set $\mathcal{T}_\mr{ChEFT}$ and the pQCD sound-speed anchor at the training density $\mathcal{T}_\mr{pQCD}=\{25\,n_0\}$, comprise our default dataset $\vec{d}_\mr{def}$.

\begin{figure*}[!t]
\includegraphics[width=\linewidth]{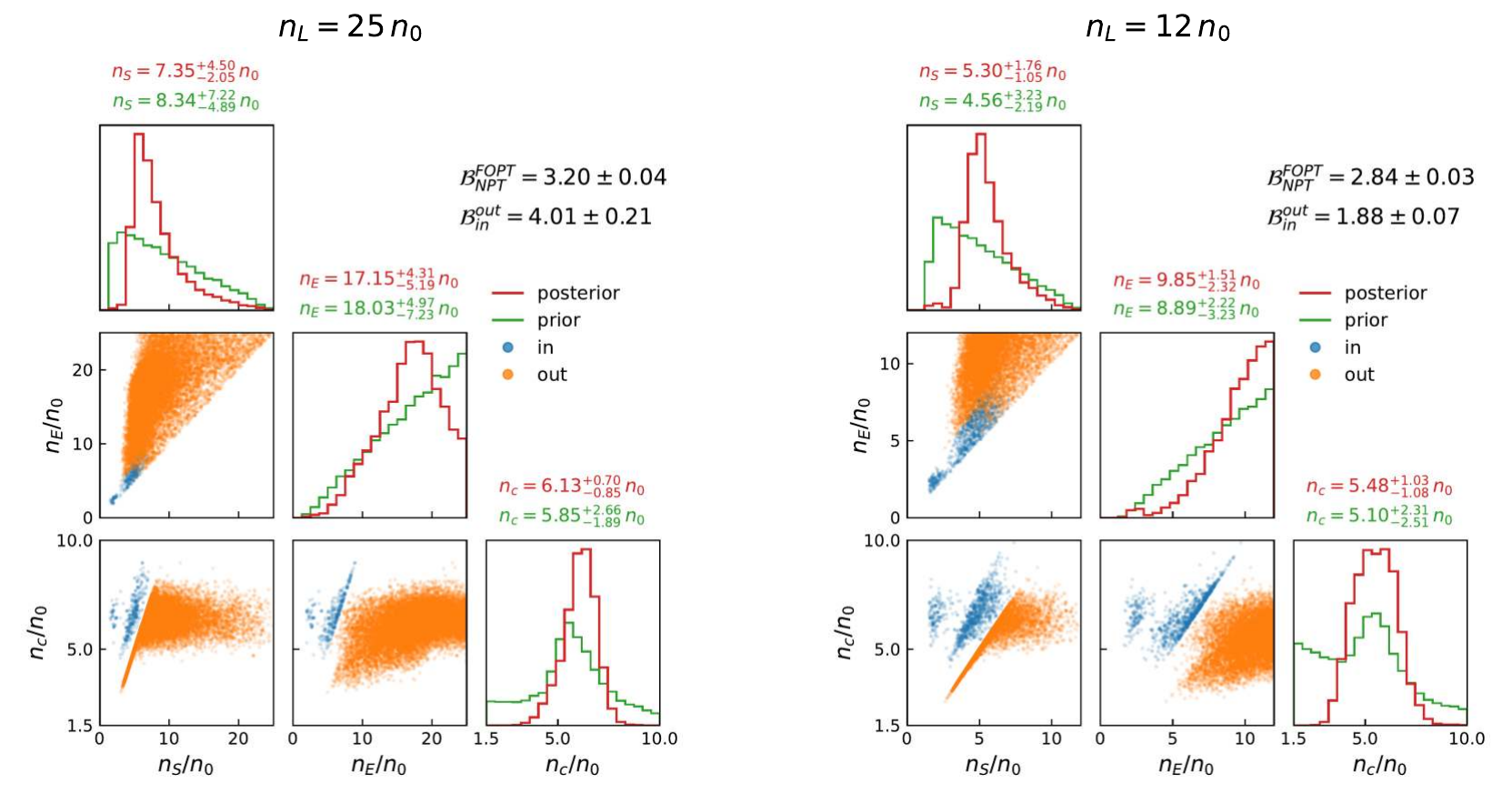}
\caption{
Corner plots of the posterior distributions of the FOPT onset density $n_S$, end density $n_E$, and central density $n_c$ of the most massive neutron star for the two choices of the termination density $n_L$: $n_L = 25\,n_0$ (left panel) and $n_L = 12\,n_0$ (right panel).
In each panel, the diagonal subpanels show the priors and one-dimensional marginalized posteriors, with the median and $68\%$ credible interval reported in each panel title; the Bayes factors $\mathcal{B}^\mr{FOPT}_\mr{NPT}$ and $\mathcal{B}^\mr{out}_\mr{in}$ are also annotated.
The off-diagonal panels display equally-weighted posterior samples, with blue ($n_S < n_c$) and orange ($n_S \ge n_c$) marking FOPT onsets below and above $n_c$, respectively.
\label{fig:nL}}
\end{figure*}

We evaluate posteriors and evidences by importance sampling.
For each hypothesis $\mathcal{H}$ and each value of $n_L$, we draw $N=8\times 10^6$ samples $\{\theta_k\}$ from the prior and assign each a weight $w_k=\prod_i\mathcal{L}(d_i\mid\theta_k,\mathcal{H})$.
The evidence then follows as $\mathcal{Z}_\mathcal{H}(\vec{d})\simeq N^{-1}\sum_k w_k$, and the effective sample size~\citep{Kong01031994}, $N_\mr{eff}=(\sum_k w_k)^2/\sum_k w_k^2$, exceeds 1 $ \times 10^4 $  in all runs, ensuring convergence of the posterior estimates.
Equally-weighted posterior samples are then obtained by resampling with replacement according to $w_k$, from which posterior distributions and Monte Carlo uncertainty are computed.

\section{Results and Discussion}
\label{sec:results}
For $n_L=25\,n_0$ with the default dataset $\vec{d}_\mr{def}$, we find $\mathcal{B}^\mr{FOPT}_\mr{NPT}=$$3.20\pm0.04$ (here and throughout, the quoted uncertainties associated with Bayes factors are Monte Carlo standard errors), corresponding to moderate evidence on the Jeffreys scale for the FOPT hypothesis relative to the NPT hypothesis.
The prior and posterior distributions of the FOPT onset density $n_S$, end density $n_E$, and the central density $n_c$ of the maximum-mass neutron star are shown in Fig.~\ref{fig:nL}, where blue and orange samples correspond to the FOPT-in ($n_S < n_c$) and FOPT-out ($n_S \ge n_c$), respectively.
Within the FOPT hypothesis, the data further favor FOPT-out over FOPT-in, with $\mathcal{B}^\mr{out}_\mr{in}=$$4.01\pm0.21$.
Together, for $n_L=25\,n_0$, the data support an FOPT whose onset most likely lies above the central density of the maximum-mass neutron star.
Placing the FOPT onset above $n_c$ agrees with earlier physics-agnostic inferences that incorporate heavy pulsars and find no sharp sound-speed drop within massive neutron-star cores~\citep{Somasundaram:2021clp,Brandes:2023hma}.
This agreement reflects a common physical tension: supporting massive neutron stars demands a stiff EOS in their interiors, while a strong FOPT would soften the EOS.

The default dataset $\vec{d}_\mr{def}$ contains two high-density pQCD inputs: the sound-speed anchor at $\mathcal{T}_\mr{pQCD}=\{25\,n_0\}$ and the marginalized pQCD likelihood.
To isolate their respective effects, each input is removed in turn at $n_L=25\,n_0$.
Removing the sound-speed anchor from $\vec{d}_\mr{def}$ while retaining the marginalized pQCD likelihood, we obtain $\mathcal{B}^\mr{FOPT}_\mr{NPT}=3.18\pm0.04$ and $\mathcal{B}^\mr{out}_\mr{in}=4.11\pm0.29$.
These values stay close to the results for $\vec{d}_\mr{def}$ ($\mathcal{B}^\mr{FOPT}_\mr{NPT}=3.20\pm0.04$ and $\mathcal{B}^\mr{out}_\mr{in}=4.01\pm0.21$), which indicates that the anchor at $\mathcal{T}_\mr{pQCD}$ has little influence on our conclusions.
Conversely, removing the marginalized pQCD likelihood from $\vec{d}_\mr{def}$ while retaining the anchor at $\mathcal{T}_\mr{pQCD}$, we find $\mathcal{B}^\mr{FOPT}_\mr{NPT}=0.79\pm0.01$ and $\mathcal{B}^\mr{out}_\mr{in}=2.53\pm0.14$, which differ substantially from the results for $\vec{d}_\mr{def}$.
The high-density pQCD constraint therefore acts predominantly through the marginalized likelihood rather than through the sound-speed anchor.

It is interesting to examine whether the astrophysical constraints and the pQCD constraint can each, on their own, drive the preference for an FOPT.
Removing both the marginalized pQCD likelihood and the sound-speed anchor at $\mathcal{T}_\mr{pQCD}$ from $\vec{d}_\mr{def}$, we obtain $\mathcal{B}^\mr{FOPT}_\mr{NPT}=0.80\pm0.01$, slightly below unity, and $\mathcal{B}^\mr{out}_\mr{in}=2.15\pm0.11$.
Removing instead the astrophysical constraints, we find $\mathcal{B}^\mr{FOPT}_\mr{NPT}=1.04\pm0.01$ and $\mathcal{B}^\mr{out}_\mr{in}=1.72\pm0.02$, both close to unity, which shows that the high-density pQCD constraint by itself yields no decisive preference.
Neither the astrophysical data nor the pQCD constraint thus provides compelling evidence for an FOPT over NPT on its own.
The support for an FOPT therefore emerges only from the interplay between the astrophysical constraints and the pQCD constraint.

In addition, to assess the sensitivity of the inferred preference for an FOPT to the prior specification, the analysis is repeated at $n_L=25\,n_0$ under three alternative choices: a log-uniform density prior and a chemical-potential prior for the FOPT location, as well as a broadened GP hyperprior.
Under the log-uniform density prior, the densities $n_S$ and $n_E$ are sampled uniformly in $\ln n$ over $1.5\,n_0\leq n\leq n_L$ and then ordered so that $n_S<n_E$.
Under the chemical-potential prior, a smooth GP EOS first defines $\mu(n)$, after which the transition chemical potential $\mu_t$ is drawn uniformly over $\mu(1.5\,n_0)\leq\mu_t\leq\mu(n_L)$ to determine $n_S$ through $\mu(n_S)=\mu_t$, and the latent heat $\Delta\varepsilon$ is drawn uniformly over $0\leq\Delta\varepsilon\leq\mu_t(n_L-n_S)$ to fix $n_E=n_S+\Delta\varepsilon/\mu_t\leq n_L$.
For the log-uniform density and chemical-potential priors, the hyperprior specified in Eq.~\eqref{eq:hyperprior} is retained, whereas for the broadened GP hyperprior, the uniform-density prior for the FOPT location is retained and all three standard deviations are doubled while their means are held fixed.
For the log-uniform density prior, the chemical-potential prior, and the broadened GP hyperprior, we find $\mathcal{B}^\mr{FOPT}_\mr{NPT}=2.24\pm0.03$, $2.65\pm0.04$, and $2.39\pm0.04$, respectively. The corresponding FOPT-out to FOPT-in Bayes factors are $\mathcal{B}^\mr{out}_\mr{in}=3.68\pm0.12$, $3.32\pm0.22$, and $3.30\pm0.22$, respectively. Thus, changing the FOPT-location prior or broadening the GP hyperprior shifts the Bayes factors only at the order-unity level, and the results still favor both the FOPT hypothesis and an onset beyond the maximum-mass central density.

For comparison, we also perform the inference under $\vec{d}_\mr{def}$ with a more restricted extrapolation boundary $n_L=12\,n_0$.
In this case, the support for the FOPT hypothesis drops to $\mathcal{B}^\mr{FOPT}_\mr{NPT}=$$2.84\pm0.03$, only anecdotal on the Jeffreys scale.
The preference for FOPT-out also weakens, from $\mathcal{B}^\mr{out}_\mr{in}=$$4.01\pm0.21$ at $n_L=25\,n_0$ to $1.88\pm0.07$ at $n_L=12\,n_0$.
Because the marginalized pQCD likelihood is built from a smooth auxiliary GP above $n_L$, lowering $n_L$ from $25\,n_0$ to $12\,n_0$ excludes any FOPT in the density range $12\,n_0<n<25\,n_0$ and thereby reduces both Bayes factors.
Even within this restricted window, nevertheless, the data still mildly favor an FOPT against NPT with the FOPT onset density lying above $n_c$.

\begin{figure}[!htbp]
\includegraphics[width=\linewidth]{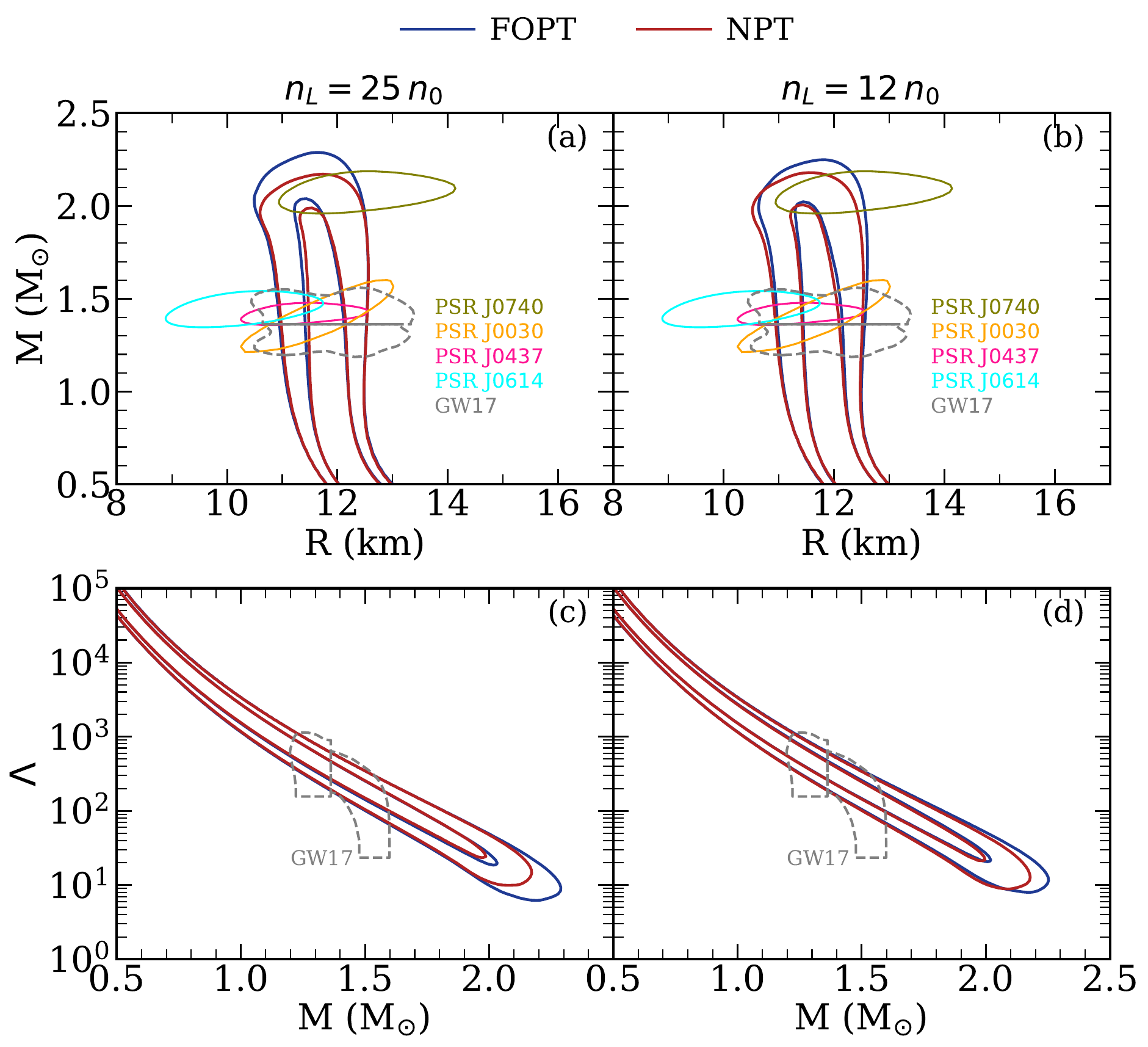}
\caption{Joint posterior distributions of mass--radius (top) and tidal deformability--mass (bottom) for neutron stars under the FOPT (blue) and NPT (red) hypotheses, shown as $68\%$ and $95\%$ credible regions.
The left (right) column is for $n_L = 25\,n_0$ ($12\,n_0$).
Astrophysical constraints---NICER joint posteriors for PSR~J0740$+$6620~\citep{Salmi:2024aum}, PSR~J0030$+$0451~\citep{Vinciguerra:2023qxq}, PSR~J0437$-$4715~\citep{Choudhury:2024xbk}, and PSR~J0614$-$3329~\citep{Mauviard:2025dmd}, together with the GW170817 posterior~\citep{LIGOScientific:2018cki}---are overlaid for reference.
\label{fig:MR}}
\end{figure}

The FOPT hypothesis shifts the maximum mass of a nonrotating neutron star $M_\mr{TOV}$ modestly upward relative to the NPT.
This shift is visible at the upper edge of the mass--radius contour in Fig.~\ref{fig:MR}(a), where $M_\mr{TOV}^\mr{NPT}=2.07^{+0.10}_{-0.08}\,M_\odot$ ($68\%$ credible interval, here and throughout unless stated otherwise) rises to $M_\mr{TOV}^\mr{FOPT}=2.15^{+0.13}_{-0.11}\,M_\odot$ for $n_L=25\,n_0$.
It should be noted that the marginalized pQCD likelihood at $n_L$ disfavors stiff EOSs and is therefore in tension with the stiffness required to support a $\sim 2\,M_\odot$ neutron star.
An FOPT resolves this tension.
The $c_s^2=0$ plateau softens the EOS and relaxes the pQCD penalty, while the inferred onset lies predominantly above $n_c$, leaving the stellar interior stiff enough for $M_\mr{TOV}$ to readily satisfy the PSR~J0740$+$6620 mass constraint.

Although the FOPT hypothesis raises $M_\mr{TOV}$, the canonical-mass observables remain essentially unchanged.
For $n_L=25\,n_0$ we obtain $R_{1.4}^\mr{FOPT}=11.76^{+0.38}_{-0.38}\,\mr{km}$ versus $R_{1.4}^\mr{NPT}=11.79^{+0.37}_{-0.41}\,\mr{km}$, and $\Lambda_{1.4}^\mr{FOPT}=309^{+88}_{-70}$ versus $\Lambda_{1.4}^\mr{NPT}=317^{+86}_{-75}$ [panel~(c) of Fig.~\ref{fig:MR}]---the two hypotheses are statistically indistinguishable at the $68\%$ level.
Earlier analyses that assume an FOPT at $2$--$3\,n_0$ soften the EOS at canonical-mass densities and drive $R_{1.4}$ down to $\sim 10\,\mr{km}$~\citep{Steiner:2012xt,Steiner:2017vmg}.
In our analysis, by contrast, $n_S$ is inferred rather than imposed, and the posterior favors an onset above the central density of the most massive neutron star, leaving the canonical-mass interior essentially unaffected.

Figure~\ref{fig:MR}(b) and~(d) show the mass--radius and tidal-deformability--mass joint posteriors for $n_L=12\,n_0$, which reproduce the qualitative conclusions established at $n_L=25\,n_0$.
The canonical-mass observables remain statistically indistinguishable between the two hypotheses at the $68\%$ level, with $R_{1.4}^\mr{FOPT}=11.79^{+0.39}_{-0.39}\,\mr{km}$ versus $R_{1.4}^\mr{NPT}=11.75^{+0.39}_{-0.39}\,\mr{km}$ and $\Lambda_{1.4}^\mr{FOPT}=314^{+92}_{-71}$ versus $\Lambda_{1.4}^\mr{NPT}=308^{+89}_{-70}$.
The modest FOPT-driven upward shift in the maximum mass also persists for $n_L=12\,n_0$, from $M_\mr{TOV}^\mr{NPT}=2.08^{+0.10}_{-0.08}\,M_\odot$ to $M_\mr{TOV}^\mr{FOPT}=2.14^{+0.12}_{-0.10}\,M_\odot$.

Because the inferred FOPT onset lies predominantly above $n_c$, the transition leaves no direct imprint on cold and stable neutron stars.
The FOPT and NPT posteriors for $R_{1.4}$ and $\Lambda_{1.4}$ agree at the $68\%$ level for either choice of $n_L$, and their $M_\mr{TOV}$ values differ by only $\Delta M_\mr{TOV}\simeq 0.06$--$0.08\,M_\odot$.
Another classical signature of an FOPT---a disconnected ``twin-star'' branch supporting two stable stellar configurations of equal mass but different radii---is likewise strongly disfavored by our analyses.
We find that the posterior probability of a twin-star solution under the FOPT hypothesis is $\le 0.1\%$ for both $n_L=25\,n_0$ and $12\,n_0$, in agreement with the model-agnostic exclusion of twin stars reported in \citet{Blomqvist:2025cxe}.

Cold and stable neutron stars thus offer no direct handle on the high-density softening induced by an FOPT, because the transition sets in beyond the densities these stars ever reach.
An FOPT remains observable, however, once attention turns to transient systems that momentarily access densities beyond the central density of the most massive stable neutron star.
Binary neutron-star post-merger gravitational-wave emission, targeted by next-generation observatories~\citep{ET:2019dnz,Ackley:2020atn,Evans:2021gyd}, offers a natural candidate: the transient remnant briefly attains densities exceeding $n_c$, and an FOPT triggered above $n_c$ would imprint itself on the gravitational-wave signal through a shift in the dominant post-merger frequency and a reduced remnant lifetime~\citep{Bauswein:2018bma,Most:2018eaw,Weih:2019xvw,Fujimoto:2022xhv}.

\begin{figure}[t]
\includegraphics[width=\linewidth]{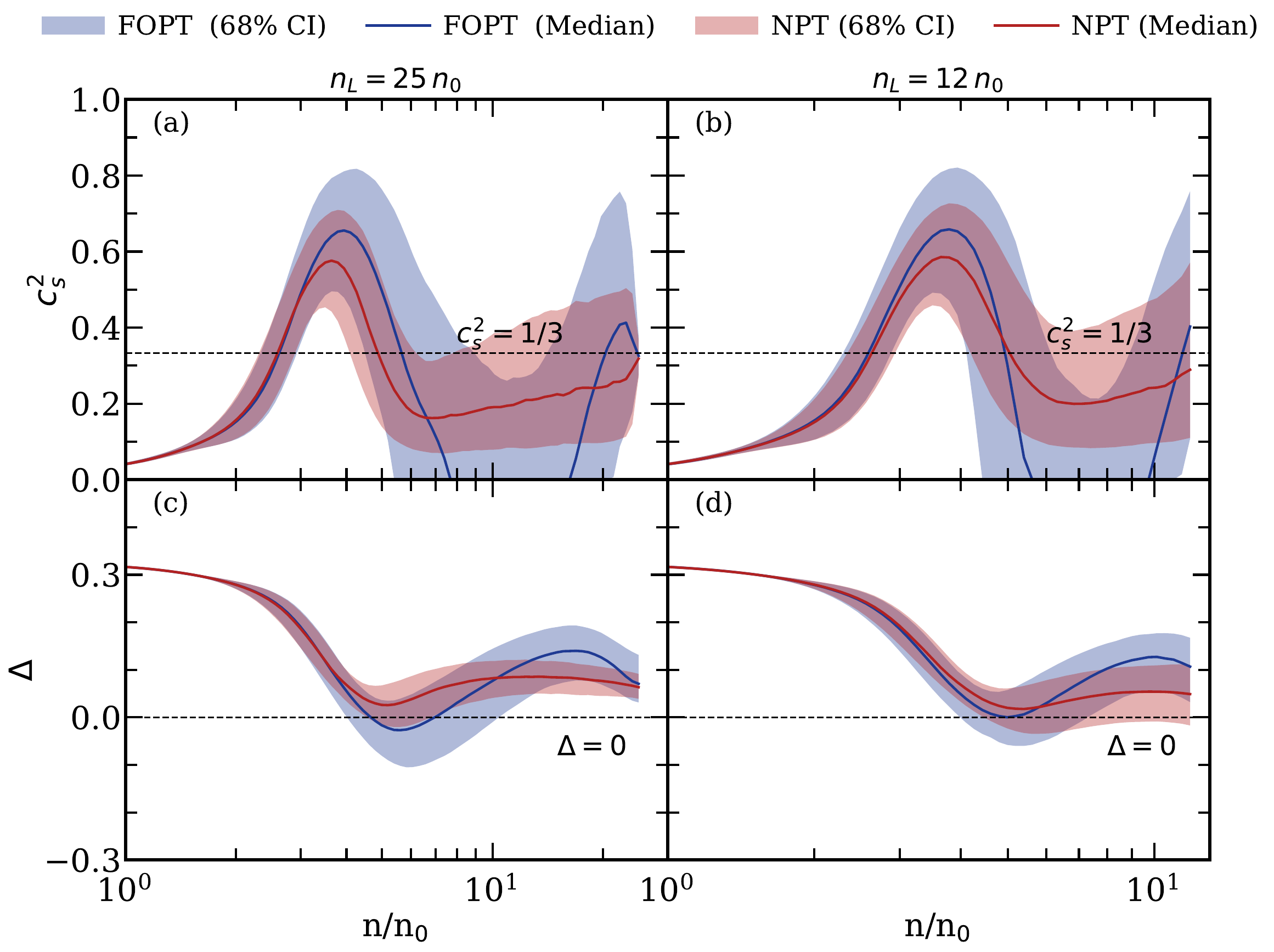}
\caption{
Squared sound speed $c_s^2$ (top) and trace anomaly $\Delta = 1/3 - p/\varepsilon$ (bottom) as functions of baryon density, with $n_L = 25\,n_0$ in the left column and $n_L = 12\,n_0$ in the right column.
Filled bands give the $68\%$ credible intervals and solid lines mark the median values, with FOPT shown in blue and NPT in red.
Black dashed lines indicate the conformal references $c_s^2 = c^2/3$ and $\Delta = 0$.
\label{fig:cs2Delta}}
\end{figure}

Figure~\ref{fig:cs2Delta}(a) shows the posterior $c_s^2(n)$ obtained with the default dataset $\vec{d}_\mr{def}$ for $n_L=25\,n_0$ under the NPT and FOPT hypotheses.
At low densities, $c_s^2$ rises under both hypotheses with essentially overlapping posteriors, consistent with the indistinguishable $R_{1.4}$ and $\Lambda_{1.4}$.
At higher densities, $c_s^2$ develops a peak near $\sim 4\,n_0$ and decreases before reaching  $n_c\simeq 6\,n_0$ in either case.
Under the FOPT, both the peak $c_s^2$ and its value at $n_c$ exceed those under the NPT, reflecting a stiffer neutron-star interior that the pQCD constraint does not overly soften.
Beyond the peak, the two posteriors approach the pQCD-predicted value ($\simeq 0.32\,c^2$) along distinct routes: the NPT descends to a minimum of $\sim 0.2\,c^2$ near $n_c$ and then rises smoothly toward the pQCD-predicted value from below, whereas the FOPT drops to zero across the plateau, rebounds to a smaller post-plateau peak, and approaches the pQCD-predicted value from above.
The post-plateau $c_s^2$ peak under the FOPT indicates that the matter beyond the FOPT remains markedly non-conformal over an extended window before joining the perturbative regime.

A complementary diagnostic of conformality is the trace anomaly $\Delta\equiv 1/3-p/\varepsilon$~\citep{Fujimoto:2022ohj}, shown in Fig.~\ref{fig:cs2Delta}(c) for $n_L=25\,n_0$.
Under the FOPT hypothesis, $\Delta$ first decreases to a dip, then rises monotonically to a pronounced maximum before approaching the perturbative regime from above, whereas the NPT posterior follows the same pattern with a markedly weaker rebound.
Moreover, $\Delta$ does not flatten toward zero below $n_c\simeq 6\,n_0$ under either hypothesis, indicating that the matter inside neutron stars remains non-conformal and strongly coupled.

\begingroup
\setlength{\tabcolsep}{3pt}
\begin{deluxetable}{lcccc}
\tablecaption{Posterior medians and $68\%$ credible intervals for the maximum nonrotating mass $M_\mr{TOV}$ ($M_\odot$), radius $R_{1.4}$ ($\mr{km}$), tidal deformability $\Lambda_{1.4}$, and central density $n_c$ of the maximum-mass configuration under the NPT and FOPT hypotheses at $n_L=25\,n_0$ and $n_L=12\,n_0$.
The FOPT onset density $n_S$, FOPT end density $n_E$, baryon-density jump $n_E-n_S$, transition pressure $p_\mr{t}$ ($\mr{MeV\,fm^{-3}}$), transition chemical potential $\mu_\mr{t}$ ($\mr{MeV}$), transition energy density $\varepsilon_\mr{t}$ ($\mr{MeV\,fm^{-3}}$), latent heat $\Delta\varepsilon$ ($\mr{MeV\,fm^{-3}}$), and dimensionless latent-heat jump $\Delta\varepsilon/\varepsilon_\mr{t}$ are reported only in the FOPT columns.
The two Bayes factors, $\mathcal{B}^\mr{FOPT}_\mr{NPT}$ (FOPT versus NPT) and $\mathcal{B}^\mr{out}_\mr{in}$ (FOPT-out versus FOPT-in), are also listed in the FOPT columns, with their quoted uncertainties given by Monte Carlo standard errors.
\label{tab:results}}
\tablehead{
\colhead{} & \multicolumn{2}{c}{$n_L=25\,n_0$} & \multicolumn{2}{c}{$n_L=12\,n_0$} \\
\cline{2-3}\cline{4-5}
\colhead{Quantity} & \colhead{NPT} & \colhead{FOPT} & \colhead{NPT} & \colhead{FOPT}
}
\startdata
$M_\mr{TOV}$ & \HPD{2.07}{0.08}{0.10}  & \HPD{2.15}{0.11}{0.13}  & \HPD{2.08}{0.08}{0.10}  & \HPD{2.14}{0.10}{0.12}  \\
$R_{1.4}$ & \HPD{11.79}{0.41}{0.37} & \HPD{11.76}{0.38}{0.38} & \HPD{11.75}{0.39}{0.39} & \HPD{11.79}{0.39}{0.39} \\
$\Lambda_{1.4}$                & \HPD{317}{75}{86}       & \HPD{309}{70}{88}       & \HPD{308}{70}{89}       & \HPD{314}{71}{92}       \\
$n_c/n_0$                      & \HPD{6.24}{0.57}{0.62}  & \HPD{6.13}{0.85}{0.70}  & \HPD{6.32}{0.57}{0.61}  & \HPD{5.48}{1.08}{1.03}  \\
$n_S/n_0$                      & ---                     & \HPD{7.35}{2.05}{4.50}  & ---                     & \HPD{5.30}{1.05}{1.76}  \\
$n_E/n_0$                      & ---                     & \HPD{17.15}{5.19}{4.31} & ---                     & \HPD{9.85}{2.32}{1.51}  \\
$(n_E-n_S)/n_0$                        & --- & \HPD{8.51}{4.42}{3.80}  & --- & \HPD{3.97}{2.28}{1.83} \\
$p_\mr{t}$                              & --- & \HPD{626}{220}{395}     & --- & \HPD{372}{105}{163} \\
$\mu_\mr{t}$                           & --- & \HPD{1857}{190}{273}    & --- & \HPD{1619}{118}{166} \\
$\varepsilon_\mr{t}$                   & --- & \HPD{1575}{559}{1316}   & --- & \HPD{999}{239}{469} \\
$\Delta\varepsilon$                    & --- & \HPD{2508}{1345}{1323} & --- & \HPD{1043}{593}{475} \\
$\Delta\varepsilon/\varepsilon_\mr{t}$   & --- & \HPD{1.56}{1.00}{1.20} & --- & \HPD{1.00}{0.64}{0.69} \\
$\mathcal{B}^\mr{FOPT}_\mr{NPT}$ & ---                    & $3.20\pm0.04$ & ---                     & $2.84\pm0.03$ \\
$\mathcal{B}^\mr{out}_\mr{in}$ & ---                     & $4.01\pm0.21$ & ---                     & $1.88\pm0.07$ \\
\enddata
\end{deluxetable}
\endgroup

Figure~\ref{fig:cs2Delta}(b) and~(d) display $c_s^2(n)$ and $\Delta(n)$ for $n_L=12\,n_0$.
Below $\sim 5\,n_0$, both panels closely follow their $n_L=25\,n_0$ counterparts: $c_s^2$ rises to a peak near $\sim 4\,n_0$ that is more pronounced under the FOPT, while $\Delta$ descends toward a dip.
At higher densities, the lower extrapolation boundary leaves a clear imprint on the FOPT plateau.
The plateau is markedly narrower than for $n_L=25\,n_0$: its median width shrinks from $n_E-n_S={8.51}_{-4.42}^{+3.80}\,n_0$ at $n_L=25\,n_0$ to ${3.97}_{-2.28}^{+1.83}\,n_0$ at $n_L=12\,n_0$, so a lower termination density compresses the density window available to the transition.
For both choices of $n_L$, Table~\ref{tab:results} summarizes the posterior medians and credible intervals of the stellar observables and characteristic densities, the thermodynamic quantities at the transition, together with the Bayes factors.

Finally, we would like to mention that the $c_s^2=0$ plateau under the FOPT follows from the Maxwell construction, in which each pure phase is separately charge-neutral and $\beta$-equilibrated.
The NPT hypothesis instead yields a smooth $c_s^2(n)$ that can in principle describe a continuous crossover from hadronic to quark matter.
In addition, an alternative treatment of the phase coexistence is the Gibbs construction, in which charge neutrality is imposed globally and the two phases coexist throughout a mixed phase~\citep{Glendenning:1992vb}.
The resulting $c_s^2(n)$ stays nonzero across the mixed phase and develops kinks at the lower and upper transition densities.
Extending our framework to the Gibbs construction is a natural next step toward a more complete characterization of the dense-matter phase transition.

\section{Summary}
\label{sec:summary}
We have performed Bayesian inference with a non-parametric Gaussian-process EOS for $\beta$-equilibrated neutron-star matter, using the GW170817 tidal deformability, the NICER mass--radius measurements of PSR~J0740$+$6620, PSR~J0030$+$0451, PSR~J0437$-$4715, and PSR~J0614$-$3329, ChEFT below $1.5\,n_0$, and the pQCD sound speed over $25\,n_0\le n\le 40\,n_0$ together with the pQCD thermodynamic quantities at $n_\mr{pQCD}\simeq 40\,n_0$.
Within this framework, we have tested the strong first-order phase transition (FOPT) hypothesis directly against the no strong first-order phase transition (NPT) hypothesis and further tested whether the FOPT onset density lies above (FOPT-out) or below (FOPT-in) the central density of the most massive neutron star.

With the Gaussian-process EOS terminated at $n_L=25\,n_0$, the data moderately favor the FOPT hypothesis over NPT and FOPT-out over FOPT-in, with $\mathcal{B}^\mr{FOPT}_\mr{NPT}=$$3.20\pm0.04$ and $\mathcal{B}^\mr{out}_\mr{in}=$$4.01\pm0.21$.
Lowering the termination density to $n_L=12\,n_0$ gives a consistent but weaker preference, with $\mathcal{B}^\mr{FOPT}_\mr{NPT}=$$2.84\pm0.03$ and $\mathcal{B}^\mr{out}_\mr{in}=$$1.88\pm0.07$.
These Bayes factors collectively favor the FOPT hypothesis, with the onset most likely \emph{beyond} the densities reached inside stable neutron stars.
Because the inferred FOPT onset predominantly lies above the central density $n_c$, the canonical-mass observables $R_{1.4}$ and $\Lambda_{1.4}$ are statistically indistinguishable between the two hypotheses.

Generally, the pQCD constraint  disfavors stiff EOSs, in tension with the stiffness required to support a $\sim 2\,M_\odot$ neutron star.
An FOPT predominantly lying above $n_c$ prevents the neutron-star interior from softening excessively under the pQCD constraint and shifts $M_\mr{TOV}$ modestly upward, while allowing the EOS to soften toward the pQCD regime at higher densities, and thus appears to provide a natural resolution of this tension.
Future gravitational-wave observations of binary neutron-star mergers, in particular the post-merger signal accessible to next-generation detectors~\citep{ET:2019dnz,Ackley:2020atn,Evans:2021gyd}, may further test our present conclusion.

\section*{Acknowledgements}
We thank Sophia Han and Zhen Zhang for helpful discussions.
This work was supported by the National Natural Science Foundation of China under Grant No. 12235010, the National SKA Program of China No. 2020SKA0120300, and the Science and Technology Commission of Shanghai Municipality (Grant No. 23JC1402700).
The computations in this paper were run on the Siyuan-1 cluster supported by the Center for High Performance Computing at Shanghai Jiao Tong University.

\bibliography{ref}
\bibliographystyle{aasjournal}

\end{document}